# Astrometric VLBI observations of $H_2O$ masers in an extreme OH/IR star candidate NSV17351

**Akiharu NAKAGAWA**[1], **Atsushi MORITA**[1], **Nobuyuki SAKAI**[2,3], **Tomoharu KURAYAMA**[4], **Hiroshi SUDOU**[5], **Gabor OROSZ**[6], **Akito YUDA**[1], **Daichi KASEDA**[1], **Masako MATSUNO**[1], **Shota HAMADA**[1], **Toshihiro OMODAKA**[1], **Yuji UENO**[7], **Katsunori M. SHIBATA**[7,8], **Yoshiaki TAMURA**[7,8], **Takaaki JIKE**[7,8], **Ken HIRANO**[7], and **Mareki HONMA**[7,8],

[1]Graduate School of Science and Engineering, Kagoshima University, 1-21-35 Korimoto, Kagoshima-shi, Kagoshima 890-0065, Japan

[2]National Astronomical Research Institute of Thailand (Public Organization), 260 Moo 4, T. Donkaew, A. Maerim, Chiang Mai, 50180, Thailand

[3]Korea Astronomy & Space Science Institute, 776, Daedeokdae-ro, Yuseong-gu, Daejeon 34055, Republic of Korea

[4]Teikyo University of Science, 2-2-1 Senju-Sakuragi, Adachi-ku, Tokyo 120-0045, Japan

[5]Gifu University, 1-1 Yanagido, Gifu City, Gifu 501-1193, Japan

[6]Joint Institute for VLBI ERIC (JIVE), Oude Hoogeveensedijk 4, 7991 PD, Dwingeloo, Netherlands

[7]Mizusawa VLBI Observatory, National Astronomical Observatory of Japan, 2-12 Hoshi-ga-oka, Mizusawa-ku, Oshu-shi, Iwate 023-0861, Japan

[8]Department of Astronomical Science, The Graduate University for Advanced Studies (SOKENDAI), Osawa 2-21-1, Mitaka, Tokyo 181-8588, Japan

∗E-mail: k1391832@kadai.jp

## Abstract

Results of astrometric very long baseline interferometry (VLBI) observations towards an ex-treme OH/IR star candidate NSV17351 are presented. We used the VERA (VLBI Exploration of Radio Astrometry) VLBI array to observe 22 GHz $H_2O$ masers of NSV17351. We derived an annual parallax of 0.247±0.035 mas which corresponds to a distance of 4.05±0.59 kpc. By averaging the proper motions of 15 maser spots, we obtained the systemic proper motion of NSV17351 to be $(\mu_a \cos \delta, \mu_\delta)^{\mathrm{avg}} = (-1.19 \pm 0.11, 1.30 \pm 0.19)$ mas yr$^{-1}$. The maser spots spread out over a region of 20 mas × 30 mas, which can be converted to a spatial distribution of ~80 au × ~120 au at the source distance. Internal motions of the maser spots suggest an outward moving maser region with respect to the estimated position of the central star. From single dish monitoring of the $H_2O$ maser emission, we estimate the pulsation period of NSV17351 to be 1122±24 days. This is the first report of the periodic activity of NSV17351, indicating that NSV17351 could have a mass of ~4 $M_\odot$. We confirmed that the time variation of $H_2O$ masers can be used as a period estimator of variable OH/IR stars. Furthermore, by inspecting dozens of double-peaked $H_2O$ maser spectra from the last 40 years, we detected a long-term acceleration in the radial velocity of the circumstellar matter to be $0.17 \pm 0.03$ km s$^{-1}$ yr$^{-1}$ Finally, we determined the position and kinematics of NSV17351 in the Milky Way Galaxy and found that NSV17351 is located in an interarm region between the Outer and Perseus arms. We note that astrometric VLBI observations towards extreme OH/IR stars are useful samples for studies of the Galactic dynamics.

**Key words:** Astrometry: — masers ($H_2O$) — stars: individual (NSV17351) — stars: variable:

## 1 Introduction

Asymptotic Giant Branch (AGB) stars are known to be at the final stage of evolution of stars with initial masses of 0.8 to 10 $M_\odot$ (e.g. Karakas & Lattanzio 2014). Among them, the stars identified as bright infrared and OH maser emitters are referred to as OH/IR stars. They represent thick circumstellar envelopes and high mass loss ratio, sometimes up to $10^{-4} M_\odot$ yr$^{-1}$ (te Lintel Hekkert et al. 1991). OH/IR stars are thought to be a group of evolved AGB stars at the stage before they evolve to planetary nebulae (te Lintel Hekkert et al. 1991; Etoka & Diamond 2006; Kamizuka et al. 2020). Same as the other types of AGB stars like Mira variables and semiregular variables, OH/IR stars often represent stellar pulsation in optical and infrared bands with typical pulsation periods of 100 to 1000 days. Engels et al. (1983) determined pulsation periods between 500 to 1800 days for 15 OH/IR stars from infrared (*K*-band) monitoring observation.

A subclass of OH/IR stars undergoing especially intensive mass loss are recognized as extreme OH/IR stars (Justtanont et al. 2013). According to the study by Höfner & Olofsson (2018), we find that sources with such high mass loss ratio have exceedingly long pulsation period, i.e., ≥800 days. Furthermore, at the late stage of AGB phase, it is known that there is also a fraction of OH/IR stars showing no or little variability, called non-variable OH/IR stars (Engels 2002). Towards bright OH/IR stars in Baud's catalog (Baud et al. 1981), Herman & Habing (1985) monitored the OH maser emission and found that 25% of the targets were non-variable OH/IR stars. In the evolution from AGB to post-AGB phase, it is thought that optical variability gradually diminishes as ceases the pulsation and heavy mass loss from the central star (e.g. Kamizuka et al. 2020).

Study of the circumstellar matter is important for understanding of the chemical properties of the Galaxy and the evolution of stars. AGB stars play a key role in the formation and transportation of circumstellar matter. OH/IR stars often host OH, $H_2O$, and SiO masers in their circumstellar envelopes (Engels 1979; Engels et al. 1986; Nyman et al. 1986). In previous research, a large amount of OH/IR stars were monitored using 1612, 1665, and 1667 MHz OH masers for determination of the OH maser flux density and its time variation (see e.g., Engels et al. 2012). Very long baseline interferometry (VLBI) observations of these masers revealed detailed structure and dynamics of circumstellar matters of AGB stars. Among them, the study by Diamond & Kemball 2003 is one of the most representative. Movies of SiO masers of TX Cam revealed ring like molecular outflows of masers explained with tangential amplification. The SiO maser shell shows significant asymmetry and can be described as a fragmented or irregular ellipsoid. Individual SiO maser components have radial motions in the range of ∼5 to 10 $km\,s^{-1}$. Decin et al. (2010) observed a nearby oxygen-rich AGB star IK Tau and presented its expansion velocity profile. The velocity data in their study were obtained from VLBI mapping studies of maser emissions from SiO, $H_2O$, and OH. The CO expansion velocity derived from ground-based CO $J = 1 - 0$ data was also considered in the study. They clarified the velocity field around an AGB star at a certain evolution phase through a wide range of radial distances, from an order of $10^{13}$ cm to $10^{16}$ cm (∼1 au to ∼1000 au). The revealed velocity profile can be an evidence for radial acceleration in the expansion velocity of the circumstellar matter. Since $H_2O$ masers occur at a radial distance of $10^{14}$ cm to $10^{15}$ cm (∼10 au to ∼100 au) where we can expect remarkable acceleration of the circumstellar envelopes, we try to explore the long-term acceleration using $H_2O$ maser data in the literature and our own observations.

NSV17351 (also named as OH224.3−1.3, IRC−10151, CRL1074, and IRAS07054−1039) is an OH/IR star (Le Squeren et al. 1979) with a spectral type of M8 (Hansen & Blanco 1975). It has OH maser emissions at 1612, 1665, and 1667 MHz (Le Squeren et al. 1979; te Lintel Hekkert et al. 1989), SiO masers at 86 GHz (Ukita & Goldsmith 1984; Engels & Heske 1989), 43 GHz (Kim et al. 2010), and H$_2$O maser at 22 GHz (Blitz & Lada 1979; Crocker & Hagen 1983; Cesaroni et al. 1988; Takaba et al. 1994; Kim et al. 2010). In a study by te Lintel Hekkert et al. (1989), a stellar LSR velocity of the source is reported to be 52 km s$^{-1}$ with no indication of its uncertainty. Accrding to a study of SiO maser by Ukita & Goldsmith (1984), a single narrow peak at 50 km s$^{-1}$ with linewidth of ∼4 km s$^{-1}$ was detected. And also in Kim et al. (2010), LSR velocities of 51.8 and 51.1 km s$^{-1}$ are presented. Based on these velocities in previous studies, it is reasonable to assume the uncertainty in the stellar LSR velocity is ∼2 km s$^{-1}$. Though the pulsation period of NSV17351 is not yet clearly given in the literature, from our observations we found the pulsation period of the source to be longer than 800 days, suggesting that NSV17351 is a candidate extreme OH/IR star.

In order to obtain physical parameters of the celestial object, distance of the source is crucial. The phase-lag method is known as a technique to derive distances of OH/IR stars (van Langevelde et al. 1990; Engels et al. 2015; Etoka et al. 2018). Distances to several OH/IR stars using the phase-lag method are reported in Engels et al. (2015). However, uncertainties of the distances from the phase-lag method are about 20 %.

Recently, the Gaia Data Release 3 (Gaia DR3; Gaia Collaboration et al. 2022)[1] provided a trigonometric parallax of 0.088±0.147 mas for NSV17351. Proper motion is also reported to be −0.03±0.16 mas yr$^{-1}$ and 1.88±0.19 mas yr$^{-1}$ in right ascension (RA) and declination (DEC), respectively. Gaia is very sensitive to extinction and size of star that is comparable to the measured parallax itself. Therefore, astrometry of OH/IR stars are considered to be essentially difficult for Gaia.

Trigonometric parallax distance measurements to a couples of long-period variables using astrometric VLBI observations have been reported (see e.g., Nakagawa et al. 2016). However, there has been a few VLBI astrometric results for OH/IR stars. A study by Orosz et al. (2017) is a notable one conducted with astrometric VLBI observations of 1612 MHz OH maser. They used the NRAO Very Long Baseline Array (VLBA)[2] and determined parallaxes of OH/IR stars. The obtained parallax of OH 138.0+7.2 was 0.52±0.09 mas, making this the first sub-mas OH maser parallax. In contrast to the compactness of H$_2$O and SiO masers, angular size of OH masers are known to be relatively extended and diffuse. OH maser parallaxes with VLBI struggle with extended maser structure and poorer resolution. Therefore, astrometric VLBI observation at higher frequency using H$_2$O masers can help us to determine smaller parallaxes with better accuracy.

Sources with pulsation periods of ∼1000 days are thought to have initial masses of ∼4 M$_\odot$

---

[1] Gaia Data Release 3; https://www.cosmos.esa.int/web/gaia/data-release-3

[2] Very Long Baseline Array; https://science.nrao.edu/facilities/vlba

(Feast 2009). Based on studies of the AGB star evolution (e.g. Vassiliadis & Wood 1993), ages of OH/IR stars with periods of 1000 days can be estimated to be ∼$10^8$ yr. Recent studies predict that galactic spiral arms are bifurcating/merging in a time scale of $10^8$ yr (Baba et al. 2013). So, OH/IR stars with ages of ∼$10^8$ yr can be used as a new probe to study the structure and evolution of spiral arms. Astrometric VLBI observations of NSV17351 is the first trial to use OH/IR stars for the studies of the Galactic dynamics.

In section 2, we give details of our VLBI observations and single dish monitoring observations, including the reduction process. In section 3, we present our results : the pulsation period, annual parallax and proper motion of NSV13751. Section 4 explains our interpretation of the $H_2O$ maser distribution and kinematics. We also discuss the evolutionary phase of NSV17351 based on the radial velocities of the $H_2O$ maser spectrum. We mention the difference of the astrometric results between VLBI and Gaia. A usefulness of extreme OH/IR stars for study of the Galactic dynamics is presented. We summarize our study in section 5 with our future prospects.

## 2 Observations and Data Reduction

### 2.1 Single dish observations

We observed $H_2O$ maser emission of NSV17351 at a rest frequency of 22.235080 GHz ($6_{16}$-$5_{23}$ transition) once every 3 to 4 weeks from August 2015 to December 2020 using the 20 m aperture telescope at VERA Iriki station in order to obtain its spectra and variability. Total number of the single dish observations is 59. Since the pulsation period of NSV17351 is not found in the literature, we estimate the pulsation period from our single dish monitoring. Integration time was 10 to 40 minutes to reduce noise levels in each observation to less than 0.05 K. The conversion factor from antenna temperature to the flux density is 19.6 Jy $K^{-1}$. A 32 MHz bandwidth data with 1024 spectral channels gives a frequency resolution of 31.25 kHz, which corresponds to a velocity resolution of 0.42 km $s^{-1}$. We carried out the data reduction using the Java NEWSTAR software package developed by the Nobeyama Radio Observatory. Amplitude of the raw spectra was calibrated by the chopper-wheel method, then the spectral baseline was corrected using a polynomial function of the seventh order. We excluded a total of 0.63 MHz signals at both ends. We adopted a signal-to-noise ratio (S/N) of 4 as a detection criterion in our single dish observations.

### 2.2 VLBI observations

We observed $H_2O$ maser emission of NSV17351 using the VLBI Exploration of Radio Astrometry (VERA). Eleven-epochs data were taken from April 2018 to June 2019 with an interval of about one month. VERA is a VLBI array which consists of four 20 m aperture radio telescopes located at Mizusawa, Iriki, Ogasawara, and Ishigaki-jima (Kobayashi et al. 2003). Its maximum baseline length is 2270 km between Mizusawa and Ishigaki-jima stations. Each antenna of VERA is equipped with a dual-beam system (Kawaguchi et al. 2000) which can simultaneously observe a target maser source and an extragalactic continuum source within a separation angle between 0.3° and 2.2°. Using the dual-beam system, we can calibrate short-term tropospheric fluctuations with the phase-referencing technique (Honma et al. 2008). Table 1 shows the nominal coordinates of the target maser source NSV17351 and extragalactic reference source J0709−1127. Regarding the revised coordinate of NSV17351 in the table, please see detail in section 4.1. Their separation angle is 0.80° at a position angle of 156°. In our phase-referencing analysis, J0709−1127 is used as a position reference on the sky plane. Dates of the VLBI observations are presented in table 2 with the Modified Julian Date (MJD). Typical integration times of the two sources were 2 to 3 hours for each VLBI observation. The signals of left-handed circular polarization from the target and position reference source were acquired with a total data recording rate of 1 giga-bit per second (Gsps). It can cover a total bandwidth of 256 MHz. The data were recorded onto the hard disk drives of the "OCTADISK" system (Oyama et al. 2016). This entire bandwidth is divided into 16 IF channels. Each IF channel then has a width of 16 MHz. Then one IF channel (16 MHz) was assigned for the maser source NSV17351 and the remaining 15 IF channels (16 MHz × 16 = 240 MHz) were assigned to the reference source J0709−1127. This process was conducted with a VERA digital filter unit (Iguchi et al. 2005). Correlation processing was done with the Mizusawa software correlator at Mizusawa VLBI observatory, NAOJ. In the final output from the correlator, the 16 MHz bandwidth data of NSV17351 was divided into 512 channels with a frequency resolution of 31.25 kHz. This corresponds to a velocity resolution of 0.42 km s$^{-1}$ at 22 GHz. In the correlator output of J0709−1127, each 16 MHz IF was divided into 32 channels.

2.3 Data reduction of the VLBI data

We reduced the VLBI data using the Astronomical Image Processing System (AIPS[1]; Greisen 2003; Fomalont 1981) developed by the National Radio Astronomy Observatory (NRAO). Amplitude calibration was performed using the gain curves and the system noise temperatures during observations at each station. A bandpass calibration was performed using the bright continuum sources DA193, OJ287, and 3C84. In the phase-referencing process, we used the task "FRING" in AIPS to solve the residual phase, group delays, and delay rates that were included in the correlator output of the reference source J0709−1127. We adopted an integration time of three minutes ("solint = 3") and solution interval of 12 seconds ("solsub = 15") in the task "FRING". The self-calibration was done using the tasks "IMAGR" and "CALIB" iteratively to solve residual phases with shorter time scale.

---

[1] AIPS; http://www.aips.nrao.edu/index.shtml

For phase calibration, we need two additional reduction procedures unique to the VERA array. Phase solutions between the dual-beam receivers, which was solved using the correlated data of noise signal injected into the two receivers from artificial noise sources installed on a feedome base of the VERA antenna (Honma et al. 2008), were also applied in the reduction process. Another calibration is related to delay-tracking models used to estimate a priori delays. Since an accuraccy of the model in the correlator is not good enough for precise astrometry, we calibrated them based on more accurate delay tracking models and applied better estimates of them. More detailed phase-referencing procedures are shown in Nakagawa et al. (2008).

Image size of the reference source is 12.8 mas × 12.8 mas square (256×256 pixels with a pixel size of 0.05 mas/pixel). The image of J0709−1127 was obtained with a peak flux density of ∼280 mJy beam$^{-1}$. Typical noise levels of the images were ∼0.9 mJy beam$^{-1}$. Then, the solutions from the tasks "FRING" and "CALIB" were transferred to the data of the target maser source NSV17351. Size of the synthesized beam was 1.7 mas × 0.9 mas with a major axis position angle of −32°. After the data calibration given above, we used the task "IMAGR" to make synthesized images of NSV17351 on 102.4 mas × 102.4 mas square maps (2048 × 2048 pixels with a pixel size of 0.05 mas/pixel). Using the task "IMFIT", we fitted two-dimensional Gaussian functions to bright maser spots to estimate their position and flux density. These positions are used in the parallax and proper motion fitting. Results of the fitting are given in section 3.2. We adopted a signal to noise ratio of 7 as a detection criterion in the phase-referenced maps.

**Table 1.** Coordinates of the sources

| Source | RA (J2000.0) | DEC (J2000.0) | $l$ | $b$ | note |
|---|---|---|---|---|---|
| NSV17351 | $07^h07^m49^s.380$ | $-10°44'05''.90$ | 224.34° | −1.29° | nominal |
|  | $07^h07^m49^s.3876 \pm 0.0004$ | $-10°44'06''.005\pm0.007$ | − | − | revised |
| J0709−1127 | $07^h09^m10^s.406578$ | $-11°27'48''.45555$ | 225.14° | −1.33° |  |

## 3 Result

3.1 Determination of the pulsation period from single dish monitoring of H$_2$O masers

We conducted 59 single dish observations of the H$_2$O maser of NSV17351 from 23 August 2015 (MJD 57257) to 8 December 2020 (MJD 59191). In the 59 observations, we detected H$_2$O maser emission in 41 observations. Figure 1 shows examples of total-power spectra of NSV17351 obtained at VERA Iriki station on 3 May 2019 (MJD 58606, top), 28 December 2018 (MJD 58480, middle), and 22 April 2018 (MJD 58230, bottom). We can see prominent maser emissions at LSR velocities ($V_{\rm LSR}$) of 39 km s$^{-1}$ and 61 km s$^{-1}$. A spectrum in 22 April 2018 (MJD 58230) represents the widest emission range in our single dish observations. A center velocity of the spectrum was obtained to be 50.1±1.9 km s$^{-1}$. The uncertainty was estimated from the full width at half maximum (FWHM) of each peak emission. This velocity is consistent with the source radial velocity of 52 km s$^{-1}$ reported by te Lintel Hekkert et al. (1989). In section 4, we use the center velocity of 50.1±1.9 km s$^{-1}$ as a representative value of a stellar LSR velocity.

**Table 2.** Dates of VLBI observations.

| Obs.ID | Year | Date | MJD |
|---|---|---|---|
| 1 | 2018 | 16 April | 58224 |
| 2 |  | 19 May | 58257 |
| 3 |  | 02 October | 58393 |
| 4 |  | 01 November | 58423 |
| 5 |  | 29 November | 58451 |
| 6 | 2019 | 04 January | 58484 |
| 7 |  | 03 February | 58517 |
| 8 |  | 12 March | 58554 |
| 9 |  | 09 April | 58582 |
| 10 |  | 04 May | 58607 |
| 11 |  | 01 June | 58635 |

In table 3, we summarize results from single dish observations at VERA Iriki station. The $T_{\rm A}^{\rm blue}$ and $T_{\rm A}^{\rm red}$ represent antenna temperatures of blue- and red-shifted velocity components in unit of K, and $V^{\rm blue}$ and $V^{\rm red}$ represent $V_{\rm LSR}$ of the components. In order to grasp overall variation of maser activity, we considered integrated intensities $I$ in unit of K km s$^{-1}$ as an integration of the whole maser components over a velocity range from 30 km s$^{-1}$ to 75 km s$^{-1}$, and presented them in 7th column in table 3. Scales of the antenna temperatures have relative uncertainties of 5-20% (Shintani et al. 2008). In this study, we uniformly applied uncertainties of 10% to all the integrated intensities. In the last column, we present rms noise levels of each single dish spectrum. Non-detections are labeled with "−" symbols.

Figure 2 shows the time variation of the integrated intensity $I$ of the H$_2$O maser of NSV17351 obtained from 23 August 2015 (MJD 57257) to 8 December 2020 (MJD 59191). Error bars indicate 10% uncertainties of integrated intensities. Horizontal axis of figure 2 represents the MJD. In the case that we could not detect any maser emission, we put open circles with downward arrows as detection upper limits of each single dish observation. From figure 2, we found that the integrated intensity $I$ of the $H_2O$ maser gradually decreased from MJD 57200 to MJD 57400, then, it further decreased bellow the detection limit (S/N of 4) of the single dish observations. The maser emission remained bellow S/N of 4 until the next detection on 08 September 2017 (MJD 58004). In March 2018 (~MJD 58200), the maser emission reached its maximum. Then it decreased and disappeared on 13 May 2019 (MJD 58616) again. NSV17351 recovered its $H_2O$ maser flux on 2 March 2020 (MJD 58910) and increased to 4.29 K km s$^{-1}$ on 8 December 2020 (MJD 59191).

Using these monitoring data, we determined the variation period of the $H_2O$ maser of NSV17351. We introduced a sinusoidal function $I_{\mathrm{model}}$ defined as follows,

$$I_{\mathrm{model}} = \Delta I \sin\left(\frac{2\pi(t+\theta)}{P}\right) + I_0, \tag{1}$$

where $\Delta I$ is the amplitude of the variation, $t$ is time, $\theta$ is a zero-phase time lag, $P$ is the variation period, and $I_0$ is an average. From our least-squares analysis, the variation period $P$ was solved to be 1122±24 days. In this fitting, we adjusted the value of $I_0$ to be 2.05. The amplitude $\Delta I$ was obtained to be 1.86 K km s$^{-1}$. The fitting solution is presented with a solid curve in figure 2. The non-detection data were not used in this fitting. Engels et al. (1986) concluded that the $H_2O$ maser luminosity of OH/IR stars follows the cycle of variation of infrared and OH luminosities, with a possible phase lag of order 0.2 relative to them. Hence, we think that the variation period of 1122±24 days estimated from our $H_2O$ maser monitoring can be considered as a pulsation period of the central star. Since our monitoring time coverage is shorter than two times of the pulsation cycle, further monitoring will be needed for careful determination of the pulsation period. Nonetheless, we note here that our estimation of the pulsation period is the first reference about the periodic activity of NSV17351and we think this source is a candidate of extreme OH/IR stars because of its long periodicity.

3.2 Annual parallax and proper motions

In this section, we determine an annual parallax of NSV17351 using positions of maser spots detected in phase-referencing analysis. We trace the angular motion of multiple maser spots on the sky plane. We succeeded to detect $H_2O$ maser spots from 1 st to 10 th observations. From 10 out of 11 VLBI observations, we found 27 $H_2O$ maser spots in 22 velocity channels. As a detection threshold, we adopted a signal-to-noise ratio (S/N) of 7. Noise levels of our phase-referenced images of NSV17351 were 80 to 170 mJy beam$^{-1}$. In the 11th observation on 01 June 2019 (MJD 58635), we were not able to detect any maser spot on the phase-referenced image. In this observation, we found maser emission in neither single dish observation nor VLBI observation. From figure 2, we can see an integrated intensity of the H$_2$O maser on 13 May 2019 (MJD 58616) decreased below our detection limit in single dish observation. The 11 th VLBI observation on 01 June 2019 (MJD 58635) was held after this non-detection.

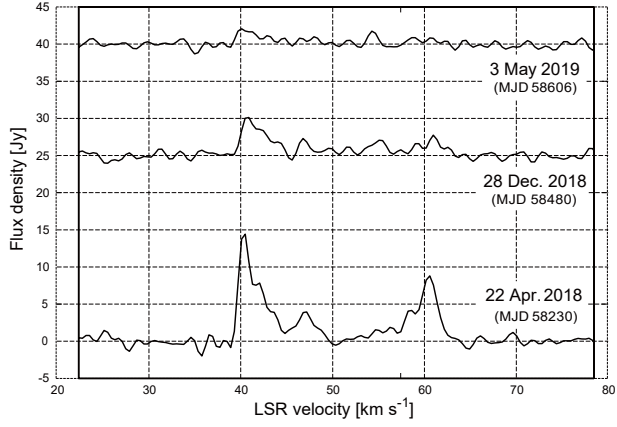

**Fig. 1.** H$_2$O maser spectra of NSV17351 obtained at VERA Iriki station on 3 May 2019 (top), 28 December 2018 (middle), and 22 April 2018 (bottom). Noise levels of individual spectra are 0.4 Jy, 0.4 Jy, and 0.8 Jy, from the top to the bottom, respectively. On 22 April 2018, the spectrum shows a double-peaked profile. For convenience, we shifted noise floors to 25 Jy and 40 Jy for spectra in middle and top, respectively.

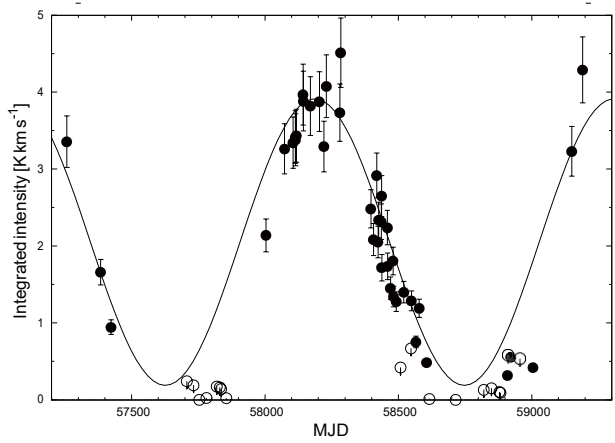

**Fig. 2.** Time variation of the integrated H$_2$O maser intensities $I$. Left and right ends correspond to 27 June 2015 (MJD 57200) and 27 March 2020 (MJD 59300), respectively. Filled circles represent results of successful detection. In the case of non-detection, we put open circles with downward arrows as representatives of detection upper limits. Solid line is the best-fit model indicating a pulsation period of 1122±24 days.

In table 4, we summarized properties of the detected maser spots. Each column represents the following quantities, maser spot ID in column 1, the LSR velocity ($V_{\rm LSR}$) in column 2, offset positions in right ascension (RA) and declination (DEC) relative to the phase tracking center in columns 3 and 4, angular proper motions in RA and DEC in columns 5 and 6, peak flux of the maser spots in column 7, signal-to-noise ratio (S/N) of the peak flux density in column 8, observation IDs where we detected maser spots in phase-referenced images in column 9. Asterisks in column 9 mean VLBI observation IDs of non-detection. When we found spatially different maser spots in an identical velocity channel, we labeled them with different spot IDs. For example, since there are two different maser spots at the $V_{\rm LSR}$ = 39.15 km s$^{-1}$, there are IDs of 3 and 4 indicating the same LSR velocity in table 4.

In figure 3, we present examples of maser spot images used in this study. From left to right, the maser spot with $V_{\rm LSR}$ of 39.15 km s$^{-1}$ (identified as ID3 in table 4) detected on (a) 16 April 2018, (b) 1 November 2018 and (c) 12 March 2019 are presented, respectively. For the spot in the map (a), formal fitted values of the peak position uncertainty are 13 $\mu$as and 22 $\mu$as in RA and DEC, respectively. For other two maps (b) and (c), we see modest structures of the maser spot. We carefully examined the maser structure, its time variation and continuity, we concluded that the southern components in the maps (b) and (c) are identical in our analysis. In the model fitting of this case, we have limited the area of its fitting to derive positions of appropriate maser components. Formal fitted values of the position uncertainty in the maps (b) and (c) are several times larger than that of the map (a). In the least-squares analysis, we regarded post-fit residuals of 0.05 mas in RA and 0.09 mas in DEC as representative errors of the maser positions across all epochs. Consequently twelve maser spots with IDs 2 to 10, 16, 23, and 24 were selected for determination of a common parallax. They were detected more than three continuous epochs of observations. Among them, maser spots with IDs of 2 and 10 were detected longer than ∼ 1 yr. The least-squares fitting gives a parallax of NSV17351 as 0.247±0.010 mas. There are some factors that contribute to the parallax uncertainty. For example, uncompensated atmospheric delay differences between the reference source and the maser would be common to all spots. Structural variation of the reference source would also be common to all spots. Therefore, in estimation of an accuracy of the parallax, we adopt a more conservative estimate. By multiplying the initial parallax error of 0.01 mas by the square-root of the number of maser spots used, we obtained 0.035 (= $0.010 \times \sqrt{12}$) mas as a true accuracy. As a result, we adopted the parallax of 0.247±0.035 mas for NSV17351 which corresponds to a distance of 4.05±0.59 kpc. Figure 4 shows offsets after removing the proper motions and the fitted parallax along RA axis (top) and DEC axis (bottom), respectively. Observed data are indicated as filled circles with their colors representing the LSR velocities of each maser spot. Error bars are 0.05 mas and 0.09 mas in RA and DEC, respectively. Solid curves are the best fit models of the parallax.

In a very recent study by Reid (2022), an imaging method with " super-resolution " was proposed. As a validation of our parallax measurement, we also performed a parallax fitting using this method. We used a round CLEAN restoring beam with 0.6 mas diameter for a maser spot showing modest structure. Using this method, a parallax was estimated to be 0.248 ± 0.035 mas showing an excellent agreement with our measurement of 0.247 ± 0.035 mas. In this paper, we adopt the latter

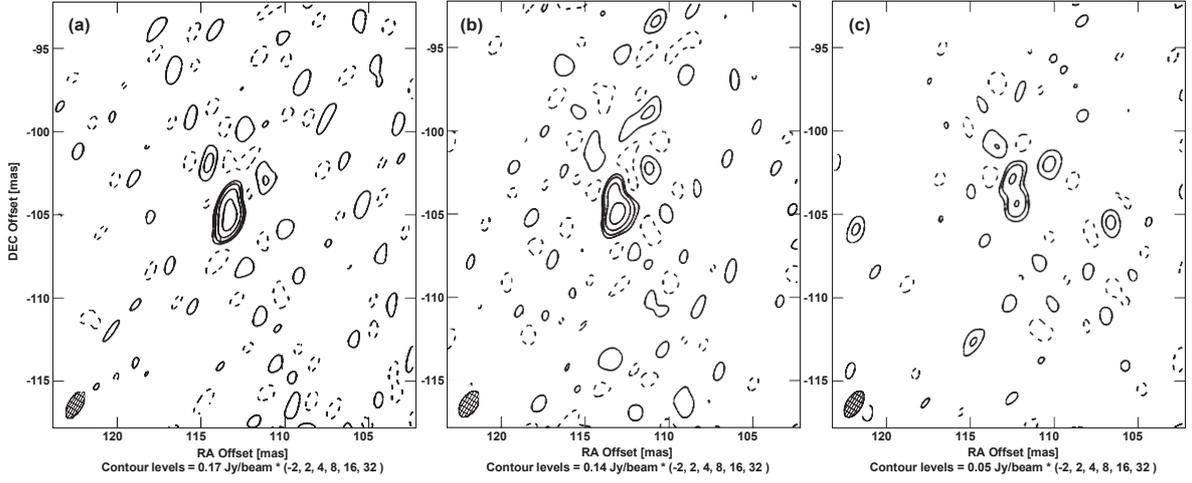

**Fig. 3.** Images of maser spots at $V_{LSR}$ of 39.15 km s$^{-1}$ detected on (a) 16 April 2018 (MJD 58224), (b) 1 November 2018 (MJD 58423) and (c) 12 March 2019 (MJD 58554). The synthesized beams are presented at bottom left of each map. In the maps of (b) and (c), southern component was used in the parallax fitting.

value as the parallax of NSV17351.

We also derived the angular proper motions of the maser spots. In the phase-referenced image, maser spots show synthesized motions of the parallax and linear proper motions. In the fitting above, we can solve the common parallax and linear proper motions of maser spots simultaneously. We successfully derived proper motions of 15 maser spots (ID 1 to 10, 16, 23, 24, 26, 27). The proper motions along the RA and DEC axes are presented in table 4 as $\mu_\alpha \cos\delta$ and $\mu_\delta$ in units of mas yr$^{-1}$, respectively. In the case when a maser spot was detected only once, or identification of the spots was difficult, we could not solve their proper motions and subsequently there are no values of proper motions. By averaging proper motions of all 15 solved maser spots, we obtained $(\mu_\alpha \cos\delta, \mu_\delta)^{avg} = (-1.19 \pm 0.11, 1.30 \pm 0.19)$ mas yr$^{-1}$ and we assume this motion as the systemic proper motion of NSV17351. Using this motion, we reveal the circumstellar motions of the maser spots in the next section.

## 4 Discussion

### 4.1 Circumstellar distribution and kinematics of the maser spots

We discuss the circumstellar kinematics of $H_2O$ maser spots of NSV17351. In figure 5, we present the distribution of all the maser spots detected in our VLBI observations. Horizontal and vertical axes of the map are position offset from the phase tracking center of NSV17351 which is given as a nominal coordinate in table 1. Position offsets of the maser spots ($\Delta\alpha \cos\delta, \Delta\delta$) are given in table 4. Maser spots are distributed in about 20 mas × 30 mas field which corresponds to ∼80 au × ∼120 au at a source distance of 4.05 kpc. From the maser distribution, we can estimate a stellar position in the map, by simply averaging positions of all the maser spots. We obtained it to be (115.49±6.35, −105.34±7.18) mas, where position uncertainties are introduced as standard deviations of all the maser spots with respect to the estimated stellar position. Estimated stellar positions are indicated by cross symbols in figure 5, the lengths of which represent the position errors on the RA and DEC axes, respectively. Based on our astromteric results, this position is presented as a revised coordinate of the source in table 1.

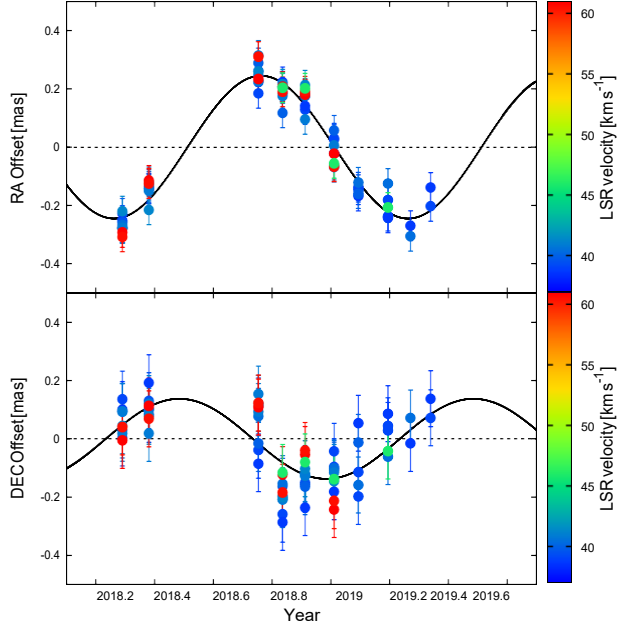

**Fig. 4.** The annual parallax of NSV17351 along RA (top) and DEC (bottom), respectively. Results from ten observations are shown with filled circles. Colors indicate LSR velocity $V_{\rm LSR}$ of each maser spot. Solid curves are the best fit models obtained from the parallax fitting.

The linear proper motions ($\mu_\alpha\cos\delta$, $\mu_\delta$) presented in the previous section derived from the phase-referencing VLBI observations are a combination of proper motion of the stellar system and internal motions of individual maser spots on a rest frame fixed to the stellar system. Therefore, to deduce their internal motions, we have to subtract the systemic motion of NSV17351 from the obtained proper motions of each maser spot. We have already derived the systemic motion of NSV17351 as $(\mu_\alpha \cos \delta, \mu_\delta)^{\rm avg} = (-1.19 \pm 0.11, 1.30 \pm 0.19)$ mas yr$^{-1}$ in previous section. If the maser positions and motions are distributed uniformly and isotropically about the star, the above systemic motion would be considered reliable values. However, it is unlikely that this condition would apply in this case, and the above systemic motion can be considered to include systematic errors. If one were to consider a realistic uncertainty in the internal motion, it would be, say, 0.5 mas yr$^{-1}$ (∼10 km s$^{-1}$). We consider adding a constant vector of about this magnitude and with a direction toward the southwest to all measured maser spot motions. In the calculations, $-0.35$ mas yr$^{-1}$ ($= -0.5/\sqrt{2}$ mas yr$^{-1}$) is added in RA and DEC. This consideration might make the distribution more likely the expansion about a central star. Taking also into account the internal motions of maser spots, the position of the central star would be considered slightly north of the position indicated by the cross symbol.

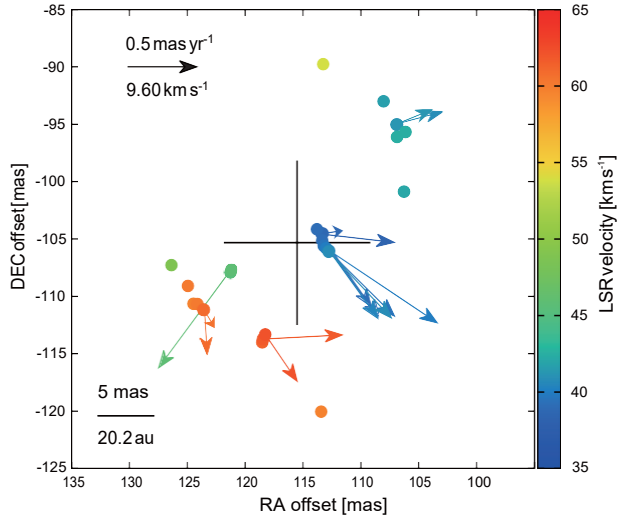

**Fig. 5.** Spatial distribution and internal motions of H$_2$O maser spots of NSV17351. Filled circles indicate maser spots and arrows indicate their internal motions. Colors indicate LSR velocities shown in a color index at right side of the map. The arrow in the top-left corner shows a proper motion of 0.5 mas yr$^{-1}$ corresponding to a velocity of 9.60 km s$^{-1}$ at 4.05 kpc. A cross symbol shows the estimated stellar position. Lengths of the cross lines indicate its errors.

In figure 5, an elongated distribution of the maser spots along the northwest to southeast direction seems to be predominant. Regarding the LSR velocity ($V_{LSR}$), we can find that blue- and red-shifted maser spots locate at northwest and southeast of the map, respectively. This indicates that these spots are likely tracing a weak, possibly asymmetric outward motion from the map center. Here we note that the OH masers observed at the position of central star are thought to be coming from the small nearside and farside parts of the envelope near the line of sight which intersects the central star (Szymczak 1988). This suggests that OH masers are pumped by infrared background radiation to which stellar photons are converted by a heavy dust shell (Orosz et al. 2017). Among all the H$_2$O maser spots in our observation, positions of the most blue-shifted maser spots are close to the estimated stellar position. Although the types of maser are different, the distribution of H$_2$O maser obtained in our observations is roughly similar to the distribution characteristics of OH maser.

We also consider the three-dimensional velocities of the maser spots. In section 3.1, we determined the stellar LSR velocity of NSV17351 to be 50.1±1.9 km s$^{-1}$. Residuals of LSR velocities ($V_{LSR}$) of each maser spot from the stellar LSR velocity of NSV17351 correspond to velocity components relative to the central star along the line of sight. Using the three orthogonal velocity components of the maser spots, we can deduce the three dimensional expanding velocity of each maser spot. The average of the expanding velocities is 15.7 km s$^{-1}$ with a standard deviation of 3.3 km s$^{-1}$.

## 4.2 Acceleration of the H$_2$O Maser

From the single dish observations at VERA Iriki station, we obtained 41 spectra of H$_2$O maser emission of NSV17351. A striking feature of the H$_2$O maser spectra in this source is the presence of blue- and red-shifted bright components with a velocity separation of ∼20 km s$^{-1}$. We can also find H$_2$O maser spectra from three previous studies. Blitz & Lada (1979), Takaba et al. (1994), and Kim et al. (2010) reported H$_2$O maser spectra observed in 28 January 1978, 10 May 1991, and 6 June 2009, respectively.

Existence of the blue- and red-shifted components seen in our observations are consistent with those reported in the previous three works, while peak velocities seem to have been slightly shifting. To argue this velocity shift, we defined $\Delta V$ as a velocity separation between the two peaks. To estimate errors of $\Delta V$, we considered the full width at half maximum (FWHM) of each component. In the studies by Blitz & Lada (1979) and Kim et al. (2010), they explicitly gave velocities of the two peaks, so we used these values to derive $\Delta V$. In the case of Takaba et al. (1994), we deduced $\Delta V$ from figure 1 in their study.

We summarized LSR velocities of the two peaks and the velocity separation $\Delta V$ in table 5. Observation dates are presented with MJD in column 1. In column 2 and column 4, peak velocities of blue- and red-shifted components are presented, respectively. In column 3 and column 5, the full width at half maximum (FWHM) values of each component are given. The velocity separation $\Delta V$ is presented in column 6 with its error obtained as an average of two FWHMs. Using all the $\Delta V$ values, we present their time variation in figure 6. From this figure, we can interpret that there is an increase in $\Delta V$ over the last 40 years. If fitting were to be performed, this increase of $\Delta V$ over a long time period would be $d\Delta V/dt = 0.33 \pm 0.06$ km s$^{-1}$ yr$^{-1}$. Here we assumed a simple linear function and the fitted was presented with solid line in figure 6. Since lifetime of individual H$_2$O maser cloud is of the order of ∼3 years (Engels 2002), it is difficult to interpret that we have been observing the same H$_2$O gas clouds during the last 40 years. Therefore, a more natural explanation is that the velocity of the region where H$_2$O masers are excited has been increasing during the last 40 years. Dividing this $d\Delta V/dt$ by two, we can obtain an acceleration of the outflow velocity to be $0.17 \pm 0.03$ km s$^{-1}$ yr$^{-1}$.

Next, we focus on the comparison of the spctrum profiles between the H$_2$O maser and OH masers. In the three H$_2$O maser spectral profiles reported in the previous studies by Blitz & Lada (1979), Takaba et al. (1994), and Kim et al. (2010), the two components have relatively gentle decreases (shoulders) at the outer sides of each peak (figure 7). On the other hand, in the spectral profiles obtained from our observation (top of figure 7), the two peaks show sharp cut offs at their outer sides. Especially, the sharpness of the cut off is remarkable at the outer side of the blue-shifted peak at 38 km s$^{-1}$ in the spectrum on 22 April 2018 (top of figure 7). Profiles of OH maser spectra are characterized by double peaked components with sharp cut offs due to a terminal velocity of the circumstellar envelope containing OH molecules. Now we can see that the profile of recently obtained H$_2$O maser spectra (top of figure 7) are quite similar to those formerly observed in 1612 MHz OH maser, which represents cut-offs at terminal velocities. In figure 8, we superposed H$_2$O maser spectrum on 22 April 2018 (solid line) and 1612 MHz OH maser spectra in February 1978 (dotted line), and we found that the profiles of the two spectra resemble each other. Especially, we can find that the cut off velocity in the blue-shifted component shows exactly same velocity (38 km s$^{-1}$ to 40 km s$^{-1}$). In the red-shifted side, velocity of the OH maser is larger than that of H$_2$O maser. It is thought that OH molecules are supplied by photodissociation of H$_2$O molecules carried to the outer part of the circumstellar envelope. This comparison indicates an asymmetric out flows of H$_2$O and OH masers in the red-shifted components.

In addition to the similarity seen in the shape of the maser spectra of H$_2$O and OH masers, we also mention the similarity of the location of the maser spots. In figure 5, the most blue-shifted H$_2$O maser spots are seen close to the estimated stellar position. In the case of OH masers, it is known that the most blue- and red-shifted maser spots are seen in the same position in the sky plane. For example, Orosz et al. (2017) revealed that the blue- and red-shifted OH masers coincide with the same position where the central star assumed to exist. In addition, we also present a study by Rosen et al. (1978). They discuss the appearance of maser emission from the gas surrounding the star by classifying limb region and far/near-side region (the direction along the line of sight) of the central star. And they reported that in the region there is rapid acceleration due to light pressure on newly formed dust grains, farside and nearside emission and limb emission are equally likely. Hence, as presumed from figure 5, we can interpret that the most blue-shifted maser spots are superposed with the position of the central star of NSV17351 along the line of sight. They can possibly be explained as the case the emission was excited along the line of sight to the central star. We mention here that there are inplane motions for the most blue-shifted maser spots.

Engels (2002) noted that the H$_2$O maser shell maintains favorable conditions for maser emission over a longer time, despite a limited lifetime for individual maser clouds on the order of ∼3 years. He suggested that H$_2$O maser clouds do not survive in the outflow but are continuously formed and destroyed. We comprehend the result of NSV17351 that the H$_2$O molecules were carried to the outermost region and the H$_2$O gas has accelerated to the terminal velocity. Since a vast amount of H$_2$O gas has been transported to the outermost regions of the circumstellar envelope, we can predict that the H$_2$O gas will soon photodissociate to OH and H, then the OH maser will brighten. The 1612 MHz OH maser line was observed with the intensity of ∼400 mJy in February 1978 (Le Squeren et al. 1979). If the OH maser emission is detected stronger than that observed in 1978 (Le Squeren et al. 1979), we might witness NSV17351 transporting H$_2$O gas to its outer region during the last 40 years. Therefore, it is important to carefully monitor OH masers of NSV17351 to study its material flow and confirm this hyopthesis.

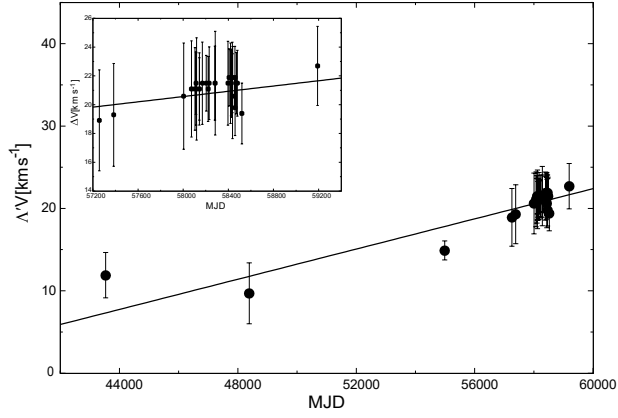

**Fig. 6.** Time variation of the velocity separation between red- and blue-shifted maser components ($\Delta V$) from 1978 to 2019 (MJD 42000 to MJD 60000). Solid line indicate a fitted model showing an acceleration of 0.33±0.06 km s$^{-1}$ yr$^{-1}$. Time variation of the $\Delta V$ between MJD 57200 and 59300 can be seen in the magnified inset at top left.

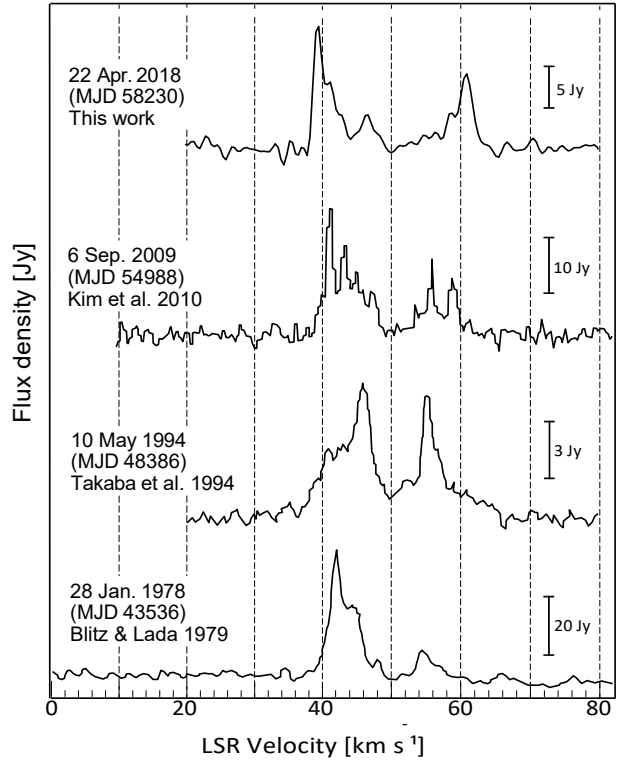

**Fig. 7.** Four representative spectra of H$_2$O maser obtained in 1978, 1994, 2009, and 2018 from bottom to top. Scales of flux density in each spectrum are presented in the right side of the figure. Time variation of the velocity profile can be seen. In the latest spectrum, we can see the largest velocity separation and sharp cut offs at the outer sides of each peak.

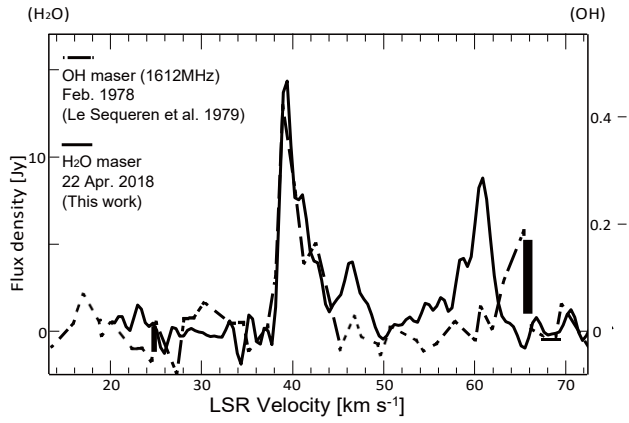

**Fig. 8.** Superpositions of H₂O maser (solid line) and OH maser (dotted line) of NSV17351 obtained in 2018 and 1978, respectively. Flux density scales in unit of Jy for H₂O and OH masers are presented in left and right vertical axes of the figure. Cut off velocity of the blue-shifted side seems to be exactly same in both spectra.

4.3 Astrometric results from VERA and Gaia

In the catalogs of Gaia Data Release 3 (DR3), we can find parallaxes of NSV17351 as 0.088±0.147 mas (Gaia Collaboration et al. 2022) which give relative errors of 170 %, while in our VLBI observations we obtained the parallax of 0.247±0.035 mas with a relative error of 14 %. The two parallaxes are barely in agreement within the error margin.

We can compare proper motions of NSV17351. The proper motions from VERA and DR3 in RA and DEC are $(\mu_\alpha \cos\delta, \mu_\delta)^{\text{avg}} = (-1.19 \pm 0.11, 1.30 \pm 0.19)$ mas yr$^{-1}$ and $(\mu_\alpha \cos\delta, \mu_\delta)^{\text{DR3}} = (-0.03\pm0.16, 1.88\pm0.19)$ mas yr$^{-1}$, respectively. Residual of the Gaia DR3 proper motions from VERA proper motions are $(\Delta\mu_\alpha \cos\delta, \Delta\mu_\delta) = (1.16 \pm 0.19, 0.58 \pm 0.27)$ mas yr$^{-1}$ which correspond to linear velocities of $(22.3\pm3.6, 11.1\pm5.2)$ km s$^{-1}$. In the study by Nakagawa et al. (2016), difference between the two measurements from VERA and HIPPARCOS was interpreted as internal motions of the maser spots. If we apply the same interpretation to NSV17351, the residual motion can be considered as internal motion of the maser spots with respect to the central star. When we assume a general property of outflow velocity of OH/IR stars or the three-dimensional outflow velocity of 15.7±3.3 km s$^{-1}$ obtained in this study (section 4.1), the velocity differences between VERA and the Gaia of $(48.6\pm7.9, -32.3\pm8.5)$ km s$^{-1}$ and $(22.3\pm3.6, 11.1\pm5.2)$ km s$^{-1}$ are large to regard them only as the internal motions of the maser spots. It should also be noted that there is a systematic uncertainty of ± 10 km s$^{-1}$ in attributing the average of the spot motions to that of the stellar system. In the comparison of the proper motions, this effect should also be considered.

## 4.4 Three-dimensional position and motion of NSV17351 in the Galaxy

We examine the three-dimensional position and motion of NSV17351 in the Galaxy. We refer to Reid et al. (2019) for transformation of the results from our VLBI observations to the position and motion in the Galactic Cartesian coordinates. We adopt the value of 50.1±1.9 km s$^{-1}$ determined in section 3.1 as the stellar LSR velocity of the source. We assumed the Galactic constants of $R_0$ = 8.15 kpc and $\Theta_0$ = 236 km s$^{-1}$, a solar motion of ($U_\odot$, $V_\odot$, $W_\odot$) = (10.6, 10.7, 7.6) km s$^{-1}$ (Reid et al. 2019), and a flat Galactic rotation curve (i.e. $\Theta(R) = \Theta_0$) in the following discussion.

We derived a three-dimensional position of NSV17351 to be ($X$, $Y$, $Z$) = (−2.83±0.12, 11.05±0.12, −0.09±0.01) kpc, where the origin of the coordinate system corresponds to the Galactic Center. From the value of $Z$ = −0.09 ± 0.01 kpc, we see that the NSV17351 is embedded in the Galactic thin disk. Since there is the offset between the physical plane and Galactic latitude b = 0 deg plane due to the Galactic warp and the tilted plane (Blaauw et al. 1960; Binney 1992), we compare the Z value of NSV17351 with the Z range of nearby star-forming regions. We confirm that the Z value of NSV17351 is included in the Z range of the SFRs (i.e., −0.12 < Z < 0.11 kpc). Figure 9 shows an enlarged view of the Milky Way Galaxy as viewed from the North Galactic pole reproduced from the Figure 2 of Reid et al. (2019). Three solid lines indicate centers of the spiral arms and grey regions indicate width of the Galactic arms enclosing 90% of sources (Reid et al. 2019). From the top to bottom, the Outer, Perseus, and Local spiral arms are shown. The filled circle indicates the position of NSV17351 with its error. Open circles indicate maser sources which have Galactocentric distances of > 7 kpc reported in a study by Reid et al. (2019). We also derived a three-dimensional noncircular motion of the source, i.e., a residual motion from the flat Galactic rotation, to be ($U, V, W$) = (−4 ± 3, −5 ± 5, −3 ± 3) km s$^{-1}$, where $U$, $V$, and $W$ are directed toward the Galactic center, the Galactic rotation, and the North Galactic pole, respectively. The errors are estimated by considering errors of the parallax, the proper motion and the systemic velocity. Details of the procedure for the error estimation are given in an appendix of Sakai et al. (2020). The obtained velocities of ($U, V, W$) are comparable to those of thin-disk sources rather than thick-disk sources that include large number of evolved stars.

In figure 9, we can find that NSV17351 is located slightly outside the Perseus arm. The distance error suggests that NSV17351 may belong to the Perseus arm, but is more likely to be in the interarm region. Indeed, if we consider the $l$-$V_{\text{LSR}}$ plot (i.e. position velocity diagram) of HI in the Figure 3 of Reid et al. (2019), we can find that the source is located on the interarm region in between the Outer arm and the Perseus arm. The location of NSV17351 in figure 9 can be understood by considering the age of the source. It is understood that pulsation period $P$ increases with increasing initial mass. Mira variable stars showing a $\log P$ of ~3.0 have initial masses of 3 to 4 $M_\odot$ (Feast 2009). By assuming this mass range, we obtained $\tau_{MS}$ of $1.6\times10^8$ to $3.5\times10^8$ years from a consideration in Sparke, & Gallagher (2000), where the $\tau_{MS}$ is main sequence life time. This indicates that the age of NSV17351 is ~$10^8$ years which is two orders of magnitude larger than the typical age of high mass star forming regions associated with spiral arms. In other words, we can say that we are observing a state that NSV17351 leaves the arm where it was born, but not yet sufficiently dynamically relaxed. Note that the spiral-arm assignment of NSV17351 should be revisited in the future because the Perseus and Outer arms are not accurately located in the Galactic 3rd quadrant due to the limited number of VLBI astrometric results.

In the last decade, VLBI astrometry has measured more than two hundred parallaxes of star forming regions (SFRs) (e.g., Burns et al. 2016; Motogi et al. 2016; Reid et al. 2019; VERA Collaboration et al. 2020) and evolved stars (e.g., Sudou et al. 2019; Kamezaki et al. 2016; Nakagawa et al. 2016; Nakagawa et al. 2018; Nakagawa et al. 2019), however ages of the sources are divided mainly into two groups. One is the age of ~$10^6$ years for SFRs, and the other is ~$10^9$ years for evolved stars. From this aspect, the extreme OH/IR star candidate NSV17351 with an estimated age of ~$10^8$ years can be regarded as a valuable source which fills the time scale gap between $10^6$ years and $10^9$ years. In recent studies of spiral arms in disk galaxies, there has been a long-standing question about how spiral arms are created and maintained. The quasi-stationary density wave theory (e.g., Lin, & Shu 1964) and the dynamic spiral theory (e.g., Sellwood, & Carlberg 1984; Baba 2015) are two major theories under discussion. Spiral arms do not show rigidly rotating patters but rather differentially rotating dynamic patterns. The amplitudes, pitch angles, and pattern speeds of spiral arms are not constant, but change within a time span of 1-2 rotational periods at each radius (Baba 2015). In the Milky Way Galaxy, rotational periods at the location of the Sun correspond to a time scale of ~$10^8$ years. For better understanding, it is important to gather samples representing various ages as suggested by previous papers (e.g., Dobbs & Pringle 2010; Miyachi et al. 2019). In this context, the extreme OH/IR stars with an age of ~$10^8$ years could be suitable samples, and astrometric VLBI is a powerful and promising method to determine their three-dimensional positions and kinematics.

## 5 Summary

We presented the first astrometric results towards an extreme OH/IR star candidate NSV17351 using the VERA VLBI array at 22 GHz. From the single dish observations, we found that NSV17351 has an extremely long period of 1122±24 days based on the variation of the $H_2O$ maser emission. From our VLBI observations, we derived an annual parallax of 0.247±0.035 mas, which corresponds to a distance of 4.05±0.59 kpc. We revealed distribution and kinematics of H$_2$O maser spots of NSV17351. Inplane distribution of 20 mas × 30 mas (∼80 au × ∼120 au at a source distance) and weak asymmetric outflow were confirmed. By averaging proper motions of maser spots, the systemic proper motion of NSV17351 was obtained to be $(\mu_\alpha \cos\delta, \mu_\delta)^{\mathrm{avg}} = (-1.19 \pm 0.11, 1.30 \pm 0.19)$ mas yr$^{-1}$. NSV17351 has a characteristics of double-peaked H$_2$O maser spectrum. We could trace the evolution of spectrums for 40 years, and we estimated the acceleration of circumstellar matter to be $0.17 \pm 0.03$ km s$^{-1}$ yr$^{-1}$.

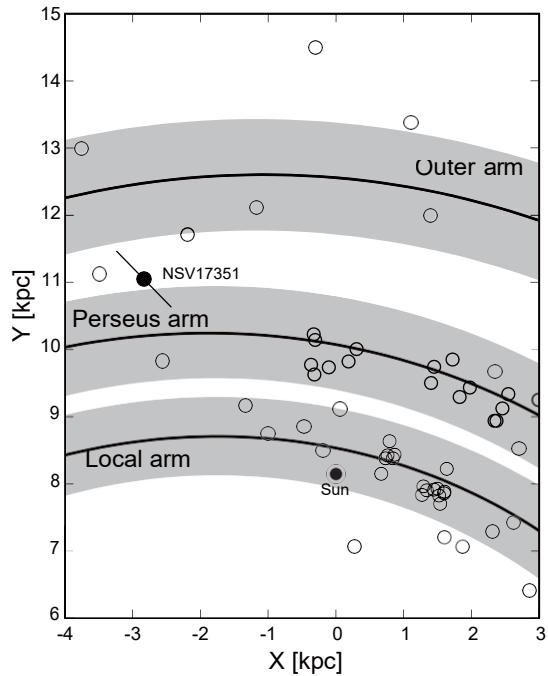

**Fig. 9.** Enlarged face-on view of the Milky Way reproduced from a study by Reid et al. (2019). The Galactic center is at (0, 0) kpc and the Sun is indicated with the symbol (⊙) at (0, 8.15) kpc. The filled circle with an error bar indicates the position of NSV17351. Open circles indicate maser sources which have Galactocentric distances of > 7 kpc in (Reid et al. 2019). Three spiral arms are presented. Solid lines indicate centers of the spiral arms. Grey regions indicate widths of the Galactic arms in which 90% of sources are enclosed (Reid et al. 2019).

We derived a three dimensional position of NSV17351 in the Milky Way Galaxy. The source was located in the interarm region between the Outer and the Perseus arms. The mass of NSV17351, inferred from its pulsation period, is 3 to 4 $M_\odot$, and its age is estimated to be ∼$10^8$ years. This is consistent with a situation that the star is located in the interam region, and away from the spiral arm where it was born.

**Acknowledgement**

We acknowledge the members of the VERA project for their kind collaboration and encouragement. Data analysis were in part carried out on common use data analysis computer system at the Astronomy Data Center, ADC, of the National Astronomical Observatory of Japan. This work has made use of data from the European Space Agency (ESA) mission *Gaia* (https://www.cosmos.esa.int/gaia), processed by the *Gaia* Data Processing and Analysis Consortium (DPAC, https://www.cosmos.esa.int/web/gaia/dpac/consortium). Funding for the DPAC has been provided by national institutions, in particular the institutions participating in the *Gaia* Multilateral Agreement. This research was supported by the leadership program of NAOJ in FY2020.

**Table 3.** Single dish observation at VERA Iriki station.

| Obs.date yyyy/mm/dd | MJD | $T_A^{\text{blue}}$ [K] | $T_A^{\text{red}}$ [K] | $V^{\text{blue}}$ [km s$^{-1}$] | $V^{\text{red}}$ [km s$^{-1}$] | $I$ [K km s$^{-1}$] | rms [K] |
|---|---|---|---|---|---|---|---|
| 2015/08/23 | 57257 | 0.46 | 0.29 | 41.09 | 60.05 | 3.35 | 0.04 |
| 2015/12/28 | 57384 | 0.16 | 0.15 | 40.90 | 60.29 | 1.66 | 0.03 |
| 2016/02/05 | 57423 | 0.15 | — | 39.88 | — | 0.94 | 0.03 |
| 2016/11/15 | 57707 | — | — | — | — | — | 0.02 |
| 2016/12/10 | 57732 | — | — | — | — | — | 0.03 |
| 2017/01/01 | 57754 | — | — | — | — | — | 0.02 |
| 2017/01/27 | 57780 | — | — | — | — | — | 0.03 |
| 2017/03/07 | 57819 | — | — | — | — | — | 0.05 |
| 2017/03/18 | 57830 | — | — | — | — | — | 0.04 |
| 2017/03/24 | 57836 | — | — | — | — | — | 0.04 |
| 2017/04/12 | 57855 | — | — | — | — | — | 0.05 |
| 2017/09/08 | 58004 | 0.22 | 0.25 | 39.39 | 60.04 | 2.14 | 0.04 |
| 2017/11/16 | 58073 | 0.49 | 0.33 | 39.58 | 60.65 | 3.26 | 0.03 |
| 2017/12/18 | 58105 | 0.48 | 0.43 | 39.39 | 60.45 | 3.34 | 0.03 |
| 2017/12/27 | 58114 | 0.53 | 0.37 | 39.25 | 60.74 | 3.41 | 0.02 |
| 2017/12/28 | 58115 | 0.53 | 0.38 | 39.40 | 60.47 | 3.38 | 0.03 |
| 2017/12/31 | 58118 | 0.55 | 0.32 | 39.62 | 60.69 | 3.43 | 0.03 |
| 2018/01/25 | 58143 | 0.64 | 0.41 | 39.67 | 60.74 | 3.97 | 0.04 |
| 2018/01/26 | 58144 | 0.61 | 0.42 | 39.58 | 60.65 | 3.88 | 0.03 |
| 2018/02/21 | 58170 | 0.68 | 0.41 | 39.40 | 60.89 | 3.82 | 0.03 |
| 2018/03/27 | 58204 | 0.69 | 0.37 | 39.82 | 61.31 | 3.88 | 0.04 |
| 2018/04/13 | 58221 | 0.69 | 0.46 | 39.40 | 60.47 | 3.29 | 0.05 |
| 2018/04/22 | 58230 | 0.74 | 0.45 | 39.40 | 60.89 | 4.08 | 0.04 |
| 2018/06/12 | 58281 | 0.78 | 0.42 | 39.41 | 60.90 | 3.73 | 0.04 |
| 2018/06/15 | 58284 | 0.70 | 0.40 | 39.40 | 60.89 | 4.51 | 0.05 |
| 2018/10/06 | 58397 | 0.39 | 0.23 | 39.22 | 60.71 | 2.48 | 0.04 |
| 2018/10/16 | 58407 | 0.44 | 0.23 | 39.23 | 61.14 | 2.08 | 0.04 |
| 2018/10/28 | 58419 | 0.37 | 0.24 | 39.41 | 60.90 | 2.92 | 0.03 |
| 2018/11/02 | 58424 | 0.40 | 0.19 | 39.39 | 60.88 | 2.05 | 0.03 |

**Table 3.** (Continued)

| Obs.date yyyy/mm/dd | MJD | $T_A^{blue}$ [K] | $T_A^{red}$ [K] | $V^{blue}$ [km s$^{-1}$] | $V^{red}$ [km s$^{-1}$] | $I$ [K km s$^{-1}$] | rms [K] |
|---|---|---|---|---|---|---|---|
| 2018/11/04 | 58426 | 0.43 | 0.20 | 39.37 | 60.86 | 2.34 | 0.03 |
| 2018/11/14 | 58436 | 0.42 | 0.20 | 39.28 | 60.76 | 2.31 | 0.03 |
| 2018/11/15 | 58437 | 0.43 | 0.15 | 39.23 | 59.88 | 2.65 | 0.04 |
| 2018/11/16 | 58438 | 0.42 | 0.22 | 39.38 | 61.29 | 1.72 | 0.04 |
| 2018/12/07 | 58459 | 0.38 | 0.16 | 38.94 | 60.85 | 2.24 | 0.04 |
| 2018/12/08 | 58460 | 0.25 | 0.24 | 39.10 | 58.91 | 1.74 | 0.05 |
| 2018/12/19 | 58471 | 0.25 | 0.20 | 39.23 | 60.72 | 1.45 | 0.04 |
| 2018/12/28 | 58480 | 0.26 | 0.14 | 39.37 | 60.86 | 1.80 | 0.02 |
| 2018/12/30 | 58482 | 0.28 | − | 38.98 | − | 1.34 | 0.05 |
| 2019/01/09 | 58492 | 0.24 | − | 40.73 | − | 1.28 | 0.04 |
| 2019/01/25 | 58508 | − | − | − | − | − | 0.04 |
| 2019/02/06 | 58520 | 0.21 | 0.12 | 39.23 | 58.61 | 1.40 | 0.02 |
| 2019/03/05 | 58547 | − | − | − | − | − | 0.04 |
| 2019/03/07 | 58549 | 0.15 | − | 39.84 | − | 1.29 | 0.02 |
| 2019/03/24 | 58566 | 0.13 | − | 38.98 | − | 0.75 | 0.02 |
| 2019/04/05 | 58578 | 0.12 | − | 38.99 | − | 1.19 | 0.02 |
| 2019/05/03 | 58606 | 0.11 | − | 38.98 | − | 0.48 | 0.02 |
| 2019/05/13 | 58616 | − | − | − | − | − | 0.03 |
| 2019/08/20 | 58715 | − | − | − | − | − | 0.04 |
| 2019/12/03 | 58820 | − | − | − | − | − | 0.03 |
| 2020/01/01 | 58849 | − | − | − | − | − | 0.02 |
| 2020/01/31 | 58879 | − | − | − | − | − | 0.02 |
| 2020/02/04 | 58883 | − | − | − | − | − | 0.02 |
| 2020/03/02 | 58910 | 0.08 | − | 38.99 | − | 0.32 | 0.02 |
| 2020/03/05 | 58913 | − | − | − | − | − | 0.03 |
| 2020/03/14 | 58922 | 0.12 | − | 39.40 | − | 0.56 | 0.03 |
| 2020/04/16 | 58955 | − | − | − | − | − | 0.02 |
| 2020/06/05 | 59005 | 0.11 | − | 39.40 | − | 0.42 | 0.02 |
| 2020/10/28 | 59150 | 0.59 | − | 38.82 | − | 3.23 | 0.05 |
| 2020/12/08 | 59191 | 0.78 | 0.24 | 38.97 | 61.73 | 4.29 | 0.02 |

**Table 4.** Kinematic properties of maser spots of NSV17351.

| Spot ID | $V_{\rm LSR}$ [km s$^{-1}$] | $\Delta\alpha\cos\delta$ [mas] | $\Delta\delta$ [mas] | $\mu_\alpha\cos\delta$ [mas yr$^{-1}$] | $\mu_\delta$ [mas yr$^{-1}$] | $S$ [Jy beam$^{-1}$] | S/N | Obs. IDs of detection |
|---|---|---|---|---|---|---|---|---|
| 1 | 38.31 | 113.37 | −104.57 | −1.41±0.21 | 1.54±0.38 | 1.03 | 11.9 | 12*456***** |
| 2† | 38.73 | 113.29 | −104.53 | −1.03±0.04 | 1.62±0.11 | 3.39 | 24.7 | 12345678910* |
| 3† | 39.15 | 113.36 | −105.11 | −1.30±0.04 | 1.04±0.11 | 5.01 | 36.4 | 12345678**** |
| 4† | 39.15 | 113.80 | −104.15 | −1.28±0.13 | 1.04±0.34 | 1.23 | 9.7 | *****678*10* |
| 5† | 39.57 | 113.22 | −105.59 | −1.40±0.07 | 1.09±0.17 | 5.27 | 29.4 | 12345****** |
| 6† | 39.99 | 113.00 | −105.82 | −1.69±0.09 | 1.06±0.10 | 1.78 | 15.6 | 1234567**** |
| 7† | 40.41 | 112.74 | −106.03 | −1.24±0.06 | 1.11±0.06 | 3.21 | 20.7 | 123456789** |
| 8† | 40.83 | 106.89 | −95.03 | −1.22±0.07 | 1.69±0.12 | 3.42 | 19.6 | 123456***** |
| 9† | 40.83 | 112.77 | −106.01 | −1.34±0.10 | 1.12±0.24 | 3.47 | 14.1 | 12345****** |
| 10† | 41.25 | 106.92 | −95.02 | −1.13±0.08 | 1.77±0.14 | 2.72 | 23.2 | 123456***** |
| 11 | 41.67 | 108.03 | −93.00 | − | − | 0.97 | 9.8 | 1***5****** |
| 12 | 42.09 | 106.26 | −100.87 | − | − | 1.40 | 12.5 | 1**4******* |
| 13 | 42.51 | 106.15 | −95.66 | − | − | 0.98 | 10.7 | 1********** |
| 14 | 42.51 | 106.89 | −96.11 | − | − | 0.71 | 7.9 | ***45****** |
| 15 | 45.45 | 121.31 | −107.97 | − | − | 0.73 | 10.1 | ****56***** |
| 16† | 45.87 | 121.24 | −107.70 | −0.37±0.22 | 0.89±0.15 | 0.96 | 9.9 | ***456*8*** |
| 17 | 48.81 | 126.36 | −107.27 | − | − | 0.83 | 7.1 | ******7**** |
| 18 | 53.85 | 113.25 | −89.76 | − | − | 0.96 | 7.2 | ******7**10* |
| 19 | 59.73 | 124.47 | −110.66 | − | − | 0.70 | 7.5 | 1********** |
| 20 | 59.73 | 113.44 | −120.03 | − | − | 0.71 | 7.6 | 1********** |
| 21 | 60.15 | 124.11 | −110.67 | − | − | 0.89 | 10.2 | 1********** |
| 22 | 60.15 | 124.95 | −109.10 | − | − | 0.49 | 7.0 | ****5****** |
| 23† | 60.57 | 123.61 | −111.18 | −0.97±0.11 | 1.47±0.16 | 2.83 | 28.4 | 123456***** |
| 24† | 60.99 | 123.55 | −111.15 | −0.91±0.07 | 1.27±0.15 | 1.79 | 17.6 | 123456***** |
| 25 | 61.41 | 118.53 | −114.01 | − | − | 2.11 | 22.6 | 12********* |
| 26 | 61.83 | 118.45 | −113.72 | −1.46±0.17 | 1.63±0.39 | 2.03 | 20.1 | 12*45****** |
| 27 | 62.25 | 118.25 | −113.32 | −1.12±0.21 | 1.26±0.16 | 1.38 | 14.8 | 12*45****** |

†: Twelve maser spots whose IDs are accompanied with daggers were used in the parallax fitting.

**Table 5.** Peak velocities of H$_2$O maser spectrum of NSV17351.

| MJD | blue [km s$^{-1}$] | | red [km s$^{-1}$] | | $\Delta V$ [km s$^{-1}$] | Reference |
|---|---|---|---|---|---|---|
| | $V_{\text{LSR}}$ | FWHM | $V_{\text{LSR}}$ | FWHM | | |
| 43536 | 42.8 | 4.0 | 54.7 | 1.5 | 11.9±2.8 | Blitz & Lada (1979) |
| 48386 | 45.6 | 5.1 | 55.3 | 2.3 | 9.7±3.7 | Takaba et al. (1994) |
| 54988 | 41.2 | 1.2 | 56.1 | 1.1 | 14.9±1.2 | Kim et al. (2010) |
| 57257 | 41.1 | 4.8 | 60.3 | 2.2 | 19.0±3.5 | this work |
| 57384 | 40.9 | 4.1 | 60.0 | 3.1 | 19.4±3.6 | this work |
| 58004 | 39.4 | 4.5 | 60.0 | 2.9 | 20.6±3.7 | this work |
| 58073 | 39.6 | 2.9 | 60.7 | 3.8 | 21.1±3.3 | this work |
| 58105 | 39.4 | 3.6 | 60.5 | 2.2 | 21.1±2.9 | this work |
| 58114 | 39.3 | 3.1 | 60.7 | 1.2 | 21.5±2.2 | this work |
| 58115 | 39.4 | 3.4 | 60.5 | 1.7 | 21.1±2.5 | this work |
| 58118 | 39.6 | 3.4 | 60.7 | 3.7 | 21.1±3.6 | this work |
| 58143 | 39.7 | 1.8 | 60.7 | 2.6 | 21.1±2.2 | this work |
| 58144 | 39.6 | 1.8 | 60.7 | 3.2 | 21.1±2.5 | this work |
| 58170 | 39.4 | 2.7 | 60.9 | 3.0 | 21.5±2.9 | this work |
| 58204 | 39.8 | 1.6 | 61.3 | 2.3 | 21.5±1.9 | this work |
| 58221 | 39.4 | 1.6 | 60.5 | 3.0 | 21.1±2.3 | this work |
| 58230 | 39.4 | 2.8 | 60.9 | 2.3 | 21.5±2.5 | this work |
| 58281 | 39.4 | 2.9 | 60.9 | 2.3 | 21.5±2.6 | this work |
| 58284 | 39.4 | 3.2 | 60.9 | 4.0 | 21.5±3.6 | this work |
| 58397 | 39.2 | 3.8 | 60.7 | 2.1 | 21.5±2.9 | this work |
| 58407 | 39.2 | 2.5 | 61.1 | 1.5 | 21.9±2.0 | this work |
| 58419 | 39.4 | 3.4 | 60.9 | 2.2 | 21.5±2.8 | this work |
| 58424 | 39.4 | 2.3 | 60.9 | 2.2 | 21.5±2.3 | this work |
| 58426 | 39.4 | 2.8 | 60.9 | 1.9 | 21.5±2.4 | this work |
| 58436 | 39.3 | 1.6 | 60.8 | 3.3 | 21.5±2.5 | this work |
| 58437 | 39.2 | 2.7 | 59.9 | 3.2 | 20.7±3.0 | this work |
| 58438 | 39.4 | 2.8 | 61.3 | 1.6 | 21.9±2.2 | this work |
| 58459 | 38.9 | 2.9 | 60.9 | 1.5 | 21.9±2.2 | this work |
| 58460 | 39.1 | 2.5 | 58.9 | 1.4 | 19.8±1.9 | this work |
| 58471 | 39.2 | 2.9 | 60.7 | 1.4 | 21.5±2.1 | this work |
| 58480 | 39.4 | 3.2 | 60.9 | 1.4 | 21.5±2.3 | this work |
| 58520 | 39.2 | 2.9 | 58.6 | 1.4 | 19.4±2.1 | this work |
| 59191 | 39.0 | 2.9 | 61.7 | 2.6 | 22.8±2.8 | this work |

| MJD | blue [km s$^{-1}$] | | red [km s$^{-1}$] | | $\Delta V$ [km s$^{-1}$] | Reference |
|---|---|---|---|---|---|---|
| | $V_{\text{LSR}}$ | FWHM | $V_{\text{LSR}}$ | FWHM | | |